\documentclass[twocolumn,showpacs,aps,prl,amsmath,amssymb,floatfix]{revtex4}
\usepackage{graphicx}
\usepackage{dcolumn}
\usepackage{bm}
\begin{document}
\title{Out-of-plane equilibrium spin current in a
quasi-two-dimensional electron gas under in-plane magnetic field}
\author{E. Nakhmedov$^{1,2}$ and O. Alekperov$^{2}$}
\affiliation{
$^1$Institut f\"ur Theoretische Physik,  Universit\"at W\"urzburg,
Am Hubland, D-97074 W\"urzburg, Germany\\
$^2$Institute of Physics, Azerbaijan National Academy of
Sciences,H. Cavid str. 33, AZ1143 Baku, Azerbaijan
}
\date{\today}
\begin{abstract}
Equilibrium spin-current is calculated in a quasi-two-dimensional 
electron gas with finite thickness under in-plane magnetic field and
in the presence of Rashba- and Dresselhaus spin-orbit interactions. The transverse confinement
is modeled by means of a parabolic potential. An orbital effect
of the in-plane magnetic field is shown to mix a transverse quantized
spin-up state with nearest-neighboring spin-down states. The out-off-plane component
of the equilibrium spin current appears to be not zero in the presence of an in-plane
magnetic field, provided at least two transverse-quantized levels are filled. 
In the absence of the magnetic field the obtained results coincide with the well-known
results, yielding cubic dependence of the equilibrium
spin current on the spin-orbit coupling constants. The persistent spin-current 
vanishes in the absence of the magnetic field if Rashba- and
Dresselhaus spin-orbit coefficients, $\alpha$ and $\beta$, are equal each other. 
In-plane magnetic field destroys this symmetry, and accumulates a finite spin-current 
as $\alpha \to \beta$.
Magnetic field is shown to change strongly the equilibrium current of the in-plane spin 
components, and gives new contributions to the cubic-dependent 
on spin-orbit constants terms. These new terms depend linearly on 
the spin-orbit constants.
\end{abstract}
\maketitle
\section{I. Introduction}
A central goal of spintronics research is an achievement of an electron spin
manipulation by means of an external electric field \cite{als02} instead of a
magnetic field, which is widely used now in semiconducting devices for
enhancement of an information processing speed. An electric field
controlled spin-orbital coupling is a promising tool \cite{dd90}
in realization of futuristic spin transport devices. During the
last ten years, there has been impressing progress in both
experimental and theoretical understanding of the spin dynamics in
quantum wells on the base of, particularly, narrow-gap
semiconductors with high g-factor, and metal-oxide-semiconductor
field-effect-transistor (MOSFET) structures. Although the
dimensionless spin-orbit (SO) coupling parameter in vacuum is as
small as $E_F/(m_0c^2)\sim 10^{-6}$, where $E_F \sim 1eV$ is the
Fermi energy of an electron and $m_0c^2 \sim 1MeV$ is the Dirac
gap, the large value of the SO coupling energy, comparable with
the Fermi energy, can be ensured by large potential gradient on
the semiconductor/insulator interface of these structures in the
presence of macroscopic structural inversion asymmetry (SIA).
Indeed, the gate potential applied across the substrate in MOSFET
results in inhomogeneous space charge distribution near the
semiconductor/insulator interface. The nonuniform macroscopic
potential, confining the electrons near the interface, varies over
a wide range, $\sim 10\div 1000~nm$, with larger potential gradient,
which originates so-called Rashba SO interaction
\cite{rashba60,br84}. On the other hand, higher value of the SO
coupling is achieved by choosing special semiconducting materials
with a bulk inversion asymmetry (BIA) in their crystalline
structure, where the gradient of the crystal potential is large.
Most prominent semiconducting compounds have either zinc-blende
structure, like GaAs and most of III-V compounds, or wurtzite
structure in II-VI compounds with BIA. Lack of the bulk inversion
symmetry in these compounds was shown by Dresselhaus
\cite{dress55} to originate another macroscopic SO interaction.

Effects of both Rashba- and Dresselhaus-SO couplings to the
physical properties of two-dimensional (2D) electron gas are not
trivial even in the absence of an external magnetic field.
Existence of a SIA or a BIA in a disordered 2D system changes
reversely the sign of the phase-coherent localization correction
to the conductivity \cite{s98,ilp94,g05}, driving the system from
a weak localization regime into an antilocalization one. Rashba-
and Dresselhaus SO interactions equally and independently
contribute to the weak antilocalization correction. Contributions
of Rashba and Dresselhaus SO interactions to the D'yakonov-Perel's
spin relaxation rate \cite{dp71} were shown to be also additive
\cite{dk86}. On the other hand, the anisotropic contribution to
the conductivity tensor \cite{cl08} in the presence of both Rashba
and Dresselhaus SO terms, the absence of spin polarization and
suppression of spin accumulation especially at the equal values of 
the coupling constants $\alpha = \pm \beta$, 
\cite{sel03,bk06,dls10}, restoration of the weak localization regime 
back at $\alpha = \pm \beta$ \cite{pp95,zfr05,skkr08} manifest an existence of the
interference between Rashba and Dresselhaus SO interactions.
Although the SO coupling generally breaks the spin rotational
symmetry, a new type of SU(2) symmetry appears \cite{bos06} in the
case of $\alpha = \beta$, which renders the spin lifetime. In the
presence of Rashba and Dresselhaus terms with equal strength, the
SO interaction rotates electron spins around a single fixed axis.
The spin along this axis becomes conserved, nevertheless spin
aligned in the perpendicular directions undergoes a deterministic
rotation depending only on the initial and final points of their
trajectory.

Different experimental techniques have been developed recently to
control a coupling of spin to the electric field
\cite{kmdg03,kmga04,lbr07}. An efficient $\hat{g}$-tensor
modulation resonance, observed in a parabolic $Al_xGa_{1-x}As$
quantum well \cite{kmdg03} with varying $Al$ content $x=x(z)$
across the well, provided an opportunity to manipulate electron
spins by means of various electron spin resonance type techniques.
An in-plane magnetic field in all of these experiments seems to be
rather favorable for getting a pronounced spin resonance.
SO interactions in a 2D electronic
system produce an effective in-plane field, which results in an
drift-driven in-plane spin polarization \cite{edelstein90}.
An external in-plane magnetic field appears to be not always summed
algebraically with SO induced effective field, and can result
in the surprising out-of-plane spin polarization \cite{erh07}, which has
been observed in a strained $n-InGaAs$ film \cite{kmga04}. On the
other hand, Hanle precession of optically oriented 2D electrons in
$GaAs$ \cite{kk90} is well described by a total in-plane
field, given as a sum of the external- and SO effective fields.
All these facts show nontrivial effects of in-plane magnetic field
on spin dynamics in quasi-2D systems. Effect of an external
magnetic field, aligned in the normal direction to the electron
gas, has been studied very well, since the problem can be solved
exactly for a non-interacting electron gas in the presence of one
of the SO interaction. In the
previous activities, the selective coupling of the in-plane
magnetic field to the electronic spin degree of freedom in the
presence of SO interactions has been used to probe the interplay
of Zeeman splitting with the SO coupling. 
The in-plane magnetic field in non-ideal 2D systems
with a finite width, which is particularly relevant for
heterojunctions and MOSFET structures, couples also to the orbital
motion, and can considerably modify the physics involved. It is
therefore important to characterize the various physical effects
generated in the presence of a parallel magnetic field, in order
to gain a better understanding of the influence of orbital
magnetic effects on the physical properties of an electron gas
with SO interactions.

Although the electron gas, formed on the semiconductor/insulator
interface in the heterojunctions and MOSFET structures, has a
finite thickness \cite{afs82}, in the most activities concerning
the SO interactions it is taken as a strictly 2D object by
neglecting the finite thickness. The thickness of the confined
electron gas in these structures is varied in the large
interval by the gate potential, applied across the electron gas, from
$10~nm$ in the inversion regime up to $1000~nm$ in the depletion
regime. The charge distribution and the electron gas thickness can
be experimentally measured and theoretically estimated with high
accuracy by means of the self-consistent solutions of
Schr\"odinger and Poisson equations under the charge balance condition. 
Finite thickness of the electron gas was recently
suggested by Rashba and Efros \cite{re03,re06} in order to study
the time-dependent gate voltage manipulation of electron spins in
MOSFETs and quantum wells, since  the spin response to a
perpendicular-to plane electric field can be achieved due to a
deviation from strict 2D limit.

In this work, we report on our investigation of both orbital and
spin effects of in-plane magnetic field in a
quasi-two-dimensional (quasi-2D) electron gas with a finite
thickness on the spin precession and splitting in the presence of
Rashba and Dresselhaus SO interactions. In this paper we calculate 
persistent spin current. Note that the model has been 
considered in our previous paper \cite{nao11a, nao11b} in order to
study the energy spectrum and the Fermi surface
under variations of SO coupling constants, the gate electric
field, the magnetic field and $g$-factor. 
Generation of a spin flux and its change under external
destructive factors is still a controversial issue
\cite{sonin08} in spintronics. Generation of a dissipationless transverse spin
current or a spin Hall current by a driving electric field ${\bf
E}$ was predicted \cite{mnz03,scns04} in a clean, infinite and
homogeneous structural inversion asymmetric 2D system. Even an
arbitrary small concentration of non-magnetic impurities was shown
\cite{ibm03,bnd04,sl04,msh04} to suppress totally the universal
value of the spin Hall conductivity peculiar to a clean system. As
it was shown by Rashba \cite{rashba03}, a spin current in the
presence of Rashba SO coupling appears even at equilibrium in
the absence of an external electric field, though it does not
result in any accumulation of spin. A universal equilibrium spin current
was shown \cite{tokatly08} to appear as a diamagnetic color current due
to a response to an effective Yang-Mills magnetic field produced by SO 
interactions, which provides an explicit realization of a non-Abelian 
Landau diamagnetism.
The equilibrium spin current in a 2D electron gas with
a slightly modulated Rashba parameter was shown \cite{sonin07} to
transfer spin from areas where spin is produced to areas where
spin  is absorbed. It was recently shown \cite{sonin07b} that an
equilibrium spin current in a 2D system with Rashba SO
interactions results in a mechanical torque on a substrate near
edge of the medium, which provides an experimental tool to detect the 
equilibrium spin current. Therefore, it can be concluded that a relation
of the equilibrium spin current to spin transport should not be
ruled out.

The central result of the paper is an appearance of out-of-plane equilibrium spin 
current in the quasi-2D electron gas under in-plane magnetic field in the presence of 
the SO interactions. In the absence of the magnetic field, the 
average values of the spin currents ${\bf J}^{S_x}$ and ${\bf J}^{S_y}$ 
are shown to coincide with the well-known results 
\cite{rashba03,esl05,bv08,sonin07} obtained for a strictly 2D electron 
gas, revealing a cubic dependence on the SO coupling constants.
We show that the magnetic field strongly changes the in-plane spin-current
components, contributing new terms
to them. The new contributions turn to be proportional, in addition
to the magnetic field, either to the gate electric field or to Zeeman splitting. 
The new terms depend linearly on the SO coupling
constants in the limiting case if one of the SO coupling constant is vanishingly small. 
Therefore, these contributions may prevail over the cubic dependent
on the SO coupling constants terms. The out-of-plane component 
${\bf J}^{S_y}=\{0,~0,~J_z^{S_z}\}$ vanishes completely with the magnetic field,
and depends quadratically or linearly on the SO coupling constants.   

The paper is organized as follows. In Section II of this work we 
describe explicitly an analytical 
solution of quantum mechanical problem of one
particle, moving in a quasi-2D system with finite thickness under
in-plane magnetic field and in the presence of Rashba and
Dresselhaus SO interactions by imposing a parabolic
confining potential in the transverse direction. 
We take into account in this work a gate potential too, 
which produces the SIA and Rashba SO
interaction. In Section III we calculate the spin current in
equilibrium. Conclusions are given in Section IV. In Appendix we
present some routine calculations of the persistent spin current.

\section{II. Energy spectrum in the presence of an in-plane magnetic field}
We consider a quasi-2D gas of electrons, moving under an external
in-plane magnetic field in the presence of both Rashba and
Dresselhaus SO interactions and a gate potential. Single particle
Hamiltonian of the system in the effective mass approximation can
be written as
 \begin{equation}
\hat{H} = \frac{{\bf P}^2}{2 m^{\ast}} + \frac{m^{\ast} \omega_0^2
z^2}{2} - e E_g z +\hat{H}_{so} + \frac{1}{2}g \mu_B {\bf \sigma}{\bf B}
\label{hamiltonian}
\end{equation}
where ${\bf P}= {\bf p} - \frac{e}{c}{\bf A}$ is an electron
momentum in the presence of a vector-potential ${\bf A}$, $m^{\ast}$
and $e$ are the electronic effective mass and charge,
respectively; $E_g$ is a strength of the gate electric field.
The second term in Eq. (\ref{hamiltonian}) is the
confining potential in $z$-direction, approximated as a parabola 
with a frequency $\omega_0$, which is a characteristic
parameter of the electron gas thickness.  This potential
does not produce a structural inversion asymmetry and,
consequently, SO interaction. Since Rashba SO interaction in the
conduction band of a semiconductor is determined by the electric
field in the valence band rather than by that in the conduction
band \cite{lassnig85}, the parabolic confinement approximation
neglects a small interface contribution to Rashba SO coupling
constant. The last term in
Eq. (\ref{hamiltonian}) is Zeeman splitting energy in the external
magnetic field ${\bf B}$ with $\omega_z \hbar = g \mu_B B/2$,
where $\mu_B = \frac{e \hbar}{2 m_0}$ is the Bohr magneton of a
free electron with mass $m_0$, $g$ is the effective Land{\'e}
factor, and ${\bf \sigma}=\{\sigma_x,\sigma_y,\sigma_z\}$ are the
Pauli spin matrices.

Spin-orbital Hamiltonian $\hat{H}_{so}$ in Eq. (\ref{hamiltonian})
contains Rashba \cite{rashba60,br84} term, $\hat{H}_R$, due to a
macroscopic SIA and Dresselhaus term \cite{dress55,pp95},
$\hat{H}_D$, due to a BIA in the crystalline structure.
Dresselhaus SO interaction in bulk semiconductors with a
zinc-blende crystal symmetry is proportional to the third order of
the electron momentum ${\bf P}$
\begin{equation}
\hat{H}_{D}=\frac{\eta}{\hbar} \sum_i \sigma_i P_i (P_{i+1}^2 -
P_{i+2}^2), (i=x, y, z;  i+3 \to i), \label{SO-D}
\end{equation}
where $\eta$ is a characteristic bulk coefficient of the SO
splitting. Since the average wave vector in the direction of the
quantum confinement $z$ is large, the terms involving $p_z^2$ will
dominate in Dresselhaus SO coupling for a quasi-2D electron gas
with finite thickness. The expression for SO interaction
Hamiltonian in MOSFETs and quantum wells of the width $d$ grown
along $[001]$ crystallographic axis reads
\begin{equation}
\hat{H}_{so}=\hat{H}_{R} + \hat{H}_{D}=\frac{\alpha}{\hbar} (\sigma_x P_y -
\sigma_y P_x) + \frac{\beta}{\hbar} (\sigma_x P_x - \sigma_y P_y),
\label{spin-orbit}
\end{equation}
where $\alpha$ and $\beta= - \eta \langle p_z^2 \rangle = - \eta
(\pi/d)^2$ are the sample dependent parameters of Rashba- and
Dresselhaus-SO interactions, correspondingly. Rashba coefficient
$\alpha$ is proportional to the gate electric field. Spin-orbital
interaction can be interpreted as an interaction of a spin with
randomly oriented, in accordance with the electron wave vector,
effective magnetic field, which lies in the plane of the electron
gas:
\begin{eqnarray}
\hat{H}_{R}&=&\frac{\hbar}{2}{\bf \sigma}\cdot {\bf \Omega}_{eff}^{R} \quad 
{\rm with} \quad {\bf
\Omega}_{eff}^{R}=\frac{2 \alpha}{{\hbar}^2}({\bf P}\times \hat
{\bf z}),\qquad
\label{so-r}\\
\hat{H}_{D}&=&\frac{\hbar}{2}{\bf\sigma}\cdot {\bf
\Omega}_{eff}^{D}, \quad {\rm with} \quad \nonumber\\
{\bf\Omega}_{eff}^{D}&=&\frac{2 \eta}{{\hbar}^2} \{P_x(P_y^2-\langle P_z^2
\rangle),P_y(\langle P_z^2 \rangle -P_x^2),0\}.\quad
\label{so-d}
\end{eqnarray}

Although the effective magnetic field
${\bf \Omega}_{eff}^R$, corresponding to Rashba term is
perpendicular to the 2D wave vector of an electron, $x$-component
of ${\bf \Omega}_{eff}^D$ is in the same direction as $p_x$ while
its $y$-component is directed in the opposite to $p_y$ direction.

An external magnetic field is chosen to be in the plane of the
2D electron gas, along $x$ axis ${\bf B}=\{B,
0, 0\}$ under the gauge ${\bf A}= \{0, -Bz, 0\}$. The magnetic
field tends to polarize the electron spin in the $x$-direction due
to the Zeeman effect, and creates an angular momentum due to the
orbital motion.

In order to solve Schr\"odinger equation $i \hbar
\frac{\partial \Psi (x,y,z;t)}{\partial t} =\hat{H}\Psi (x,y,z;t)$
with a spinor $\Psi (x,y,z;t)= {\Psi_{\uparrow} \choose
\Psi_{\downarrow}}$ one expresses the electron wave functions with
spin-up $\Psi_{\uparrow}$ and spin-down $\Psi_{\downarrow}$ orientations as
$\Psi_{\uparrow,\downarrow}(x,y,z;t) = e^{ik_x x +ik_y
y}\psi_{\uparrow,\downarrow}(z,t)$ which yields
\nopagebreak
\begin{widetext}
\begin{eqnarray}
i\hbar \frac{\partial\psi_{\uparrow}}{\partial t} =
\left\{-\frac{\hbar^2}{2 m^{\ast}} \frac{\partial^2}{\partial z^2}
+ \frac{m^{\ast} \omega^2 z^2}{2} + (k_y \hbar \omega_B - e E_g) z +
\frac{\hbar^2 k^2}{2 m^{\ast}} \right\} \psi_{\uparrow} +
\big[(\alpha + i \beta)(k_y+\omega_B m^{\ast} z/\hbar) +
(\beta+i\alpha) k_x + \hbar \omega_z \big]
\psi_{\downarrow};
\label{eq-up}\\
i \hbar \frac{\partial \psi_{\downarrow}}{\partial t} =
\left\{-\frac{\hbar^2}{2 m^{\ast}} \frac{\partial^2}{\partial z^2}
+ \frac{m^{\ast} \omega^2 z^2}{2} + (k_y \hbar \omega_B - e E_g) z +
\frac{\hbar^2 k^2}{2 m^{\ast}}\right\} \psi_{\downarrow} +
\big[(\alpha - i \beta)(k_y + \omega_B m^{\ast} z/\hbar) + (\beta
- i \alpha)k_x +\hbar \omega_z \big] \psi_{\uparrow},
\label{eq-down}
\end{eqnarray}
\end{widetext}
where $\omega_B=eB/m^{\ast} c$ is the cyclotron frequency, $\omega
= \sqrt {\omega_B^2 + \omega_0^2}$ is the effective frequency, and
$k = \sqrt {k_x^2 + k_y^2}$ is the modulus of a 2D-wave vector. In
the stationary case the wave function is chosen as $\Psi
(x,y,z;t)= exp(-iEt/\hbar)\Psi(x,y,z)$, where $E$ is the total
energy of an electron. In the absence of SO interactions and
Zeeman term, Eqs. (\ref{eq-up}) and (\ref{eq-down}) are decoupled
and reduced to the oscillator equation with real wave function
\begin{equation}
\left\{-\frac{\hbar^2}{2 m^{\ast}}\frac{d^2}{d z^2} +
\frac{m^{\ast} \omega^2}{2} (z - z_0)^2  - \tilde{E}\right\} \psi^{(0)}(z) = 0,
\label{hermite}
\end{equation}
where $z_0$ is the z-coordinate of a magnetic orbit,
$z_0 = \frac{e E_g - k_y \hbar \omega_B}{m^{\ast} \omega^2}$,
and $\tilde{E} = E-\frac{\hbar^2 k^2}{2
m^{\ast}} + \frac{(e E_g - \hbar k_y \omega_B)^2}{2
m^{\ast}\omega^2}$ is the energy spectrum of
the quantized orbital, $\tilde{E}_n = \hbar \omega(n+ 1/2)$,
corresponding to the $n$th state
\begin{equation}
\psi_n^{(0)}(z)=(\sqrt \pi 2^n n!)^{-1/2}
e^{-\frac{(z - z_0)^2}{2a_B^2}} H_n\left(\frac{z - z_0}{a_B}\right)
\label{zero-psi}
\end{equation}
with $H_n(z)$ and $a_B=\sqrt{\hbar/m^{\ast}\omega}$ being the
Hermite polynomial and the Bohr radius, correspondingly.

SO interactions in Eqs. (\ref{eq-up}) and (\ref{eq-down}) mix the transverse-quantized levels,
yielding the complex wave functions  
$\psi_{\uparrow}(z)$ and $\psi_{\downarrow}(z)$; they 
furthermore satisfy the condition $\psi_{\uparrow, \downarrow} =
e^{i\theta} \psi^{\ast}_{\downarrow, \uparrow}$. A coordinate-dependent term, 
$\propto \omega_B m^{\ast}z$, in the off-diagonal part of 
Eqs. (\ref{eq-up}) and (\ref{eq-down}) is originated from the orbital magnetic field  
effect, which links $n$th orbital of a spin-up electron with 
$(n \pm 1)$th orbital of a spin-down electron and vice versa. 
Eqs. (\ref{eq-up}) and (\ref{eq-down}) are easily solved in the absence of this
spatial-dependent term. Indeed, let us replace $z$ in the 'mixing' terms of
Eqs. (\ref{eq-up}) and (\ref{eq-down}) by the coordinate of the
magnetic orbital center $z_0$, and seek the solution as
$\Psi_n^{(0)} (x,y,z) = e^{ik_x x + i k_y y} \psi_n^{(0)}(z){A_n
\choose B_n}$. We get the following system of equations for $A_n$
and $B_n$
\begin{eqnarray}
(\tilde{E}/\hbar \omega - n - 1/2)
A_n -c_0 B_n = 0\nonumber\\
-c_0^{\ast} A_n + (\tilde{E}/\hbar \omega - n - 1/2) B_n = 0,
\label{zero-coeff}
\end{eqnarray}
where the dimensionless coefficient $c_0$ is given
\begin{equation}
c_0=\frac{1}{\hbar \omega}\left[(\alpha + i\beta)\left(k_y
\frac{\omega_0^2}{\omega^2} + \frac{e E_g \omega_B}{\hbar \omega^2}\right) + (i\alpha + \beta) k_x +
\omega_z \hbar \right]. \label{c0}
\end{equation}
The energy spectrum is immediately obtained from
Eq.(\ref{zero-coeff})
\begin{widetext}
\begin{eqnarray}
E_n^{\pm}(k_x,k_y) = \frac{\hbar^2 k^2}{2m^{\ast}} -
\frac{(k_y \hbar \omega_B - e E_g)^2}{2  m^{\ast} \omega^2} +
\omega \hbar (n+ 1/2) \pm 
\Bigg\{(\alpha^2 +
\beta^2) \left[ k_x^2 + \frac{(k_y \omega_0^2 + e E_g \omega_B/\hbar)^2}{\omega^4}\right] +\nonumber\\
+ 4\alpha \beta k_x \frac{(k_y \omega_0^2 + e E_g \omega_B/\hbar)}{\omega^2} + \omega_z^2\hbar^2 + 2
\omega_z \hbar \left[\alpha \frac{(k_y \omega_0^2 + e E_g \omega_B/\hbar)}{\omega^2} + \beta k_x\right]
\Bigg\}^{1/2}. \label{approx-energy}
\end{eqnarray}
\end{widetext}
The coefficients $A_n$ and $B_n$ in the spinor are completely
defined from the normalization condition $|A_n|^2 + |B_n|^2=1$ and
Eq. (\ref{zero-coeff})
\begin{equation}
A_n = \frac{1}{\sqrt 2} \qquad  {\rm and}  \qquad  B_n = \pm
\frac{|c_0|}{{\sqrt 2}~ c_0}.
\end{equation}

General solutions of Eqs. (\ref{eq-up}) and (\ref{eq-down}) are
sought as linear combinations of $\psi_n^{(0)}(z)$
\begin{eqnarray}
\psi_{\uparrow}(z) =
e^{-\frac{(z - z_0)^2}{2a_B^2}}\sum_{n=0}^{\infty}\frac{a_n}{\sqrt{a_B
\sqrt \pi 2^n n!}} H_n\left(\frac{z - z_0}{a_B}\right);\quad
\label{wave-up}\\
\psi_{\downarrow}(z) =
e^{-\frac{(z - z_0)^2}{2a_B^2}}\sum_{n=0}^{\infty}\frac{b_n}{\sqrt{a_B
\sqrt \pi 2^n n!}} H_n\left(\frac{z - z_0}{a_B}\right),\quad
\label{wave-down}
\end{eqnarray}
where the coefficients $a_n$ and $b_n$ satisfy the
normalization condition $\sum_{n=0}^{\infty} (|a_n|^2 +
|b_n|^2)=1$. From the condition that Eqs. (\ref{eq-up}) and
(\ref{eq-down}) are complex conjugate each other, one gets
$a_n=e^{i\theta} b_n^{\ast}$ and $b_n = e^{i\theta} a_{n}^{\ast}$
with $\theta$ being a real phase shift. Therefore, $\sum_n |a_n|^2
= \sum_n |b_n|^2 = \frac{1}{2}$. It is easy to estimate the
average value of the spin operator components ${\bf S} =
\{\hat{S}_x,\hat{S}_y,\hat{S}_z\} = \hbar/2
\{\sigma_x,\sigma_y,\sigma_z\}$ over the stationary states given
by Eqs. (\ref{wave-up}) and (\ref{wave-down})
\begin{eqnarray}
\langle \hat{S}_x \rangle_{\bf k} = \frac{\hbar}{2} \sum_n(a_n^{\ast} b_n
+ b_n^{\ast} a_n),\qquad \qquad \qquad \qquad \qquad \nonumber\\
\langle \hat{S}_y \rangle_{\bf k} = -i \frac{\hbar}{2} \sum_n(a_n^{\ast}
b_n - b_n^{\ast} a_n),\qquad \qquad \qquad \qquad \nonumber\\
\langle \hat{S}_z \rangle_{\bf k} = \frac{\hbar}{2} \sum_n(|a_n|^2 -
|b_n|^2) = 0,\qquad \qquad \qquad \qquad \nonumber\\
\langle \hat{S}^{+} \rangle_{\bf k} = \hbar \sum_n a_n^{\ast} b_n, \quad
{\rm and} \quad \langle \hat{S}^{-} \rangle_{\bf k} = \hbar \sum_n
b_n^{\ast} a_n. \qquad
\end{eqnarray}
So a spin precesses around the normal to the plane, and $\langle
\hat{S}_z\rangle_{\bf k}$ averages out to zero, whereas in-plane
components of the spin take finite values. The average position of
an electron in the confining potential $\langle z \rangle$ can be
calculated by the same way,
\begin{eqnarray}
\langle z \rangle_{\bf k} = \frac{a_B}{\sqrt{2}} \sum_n  \sqrt{n+1}~ \{(
a_{n+1}^{\ast} a_n + a_n^{\ast} a_{n+1}) + \nonumber\\
(b_{n+1}^{\ast} b_n + b_n^{\ast} b_{n+1}) \} + z_0, \label{center}
\end{eqnarray}
which means that an overlap between the
neighboring transverse-quantized levels shifts the center of the magnetic orbit
of both spin-up and spin-down electrons equally, in addition to
the magnetic- and gate electric fields shift $z_0$, in
$z$-direction. 

Equations for the coefficients $a_n$ and $b_n$ with $n=
0,1,2,3,\dots$ can be obtained by putting Eqs. (\ref{wave-up}) 
and (\ref{wave-down}) into Eqs. (\ref{eq-up}) and (\ref{eq-down})
\begin{eqnarray}
\left(\frac{\tilde {E}}{\hbar \omega} - n - \frac{1}{2}\right) a_n
-c_0 b_n - \sqrt{2n}~ c_1~ b_{n-1} -\nonumber\\
-\sqrt{2(n+1)}~ c_1~ b_{n+1} = 0,
\label{eq-a}\\
\left(\frac{\tilde {E}}{\hbar \omega} - n - \frac{1}{2}\right) b_n
-c_0^{\ast} a_n - \sqrt{2n}~ c_1^{\ast}~ a_{n-1} -
\nonumber\\
-\sqrt{2(n+1)}~ c_1^{\ast}~ a_{n+1} = 0, \label{eq-b}
\end{eqnarray}
where $c_0$ is defined by Eq. (\ref{c0}), and the coefficient $c_1$
is given as
\begin{equation}
c_1 = (\alpha + i \beta)\frac{\omega_B}{2\hbar
\omega}\sqrt{\frac{m^{\ast}}{\omega \hbar}}. \label{c1}
\end{equation}
Note that $a_n=b_n=0$ for $n<0$ in Eqs. (\ref{eq-a}) and (\ref{eq-b}).

It is easy to see that the approximate equations
(\ref{zero-coeff}) can be obtained from Eqs.(\ref{eq-a}) and
(\ref{eq-b}) by neglecting all terms $\sim c_1$.

In the absence of the external magnetic field, $B=0$, the
expressions for $c_0$ and $c_1$, given by Eqs.(\ref{c0}) and
(\ref{c1}), are simplified
\begin{equation}
c_0 = \frac{1}{\omega_0 \hbar} [i \alpha (k_x - i k_y) + \beta (k_x + i
k_y)], \quad {\rm and} \quad c_1 = 0,
\end{equation}
and, as a result, a mixing between the transverse-quantized levels is left
off (see, Eqs. (\ref{eq-a}) and
(\ref{eq-b})). A simple exact expression for the energy spectrum in
the absence of the magnetic field is obtained
\begin{eqnarray}
E_n^{\pm} = \hbar \omega_0 (n + \frac{1}{2}) + \frac{\hbar^2
k^2}{2 m^{\ast}} - \frac{e^2E_g^2}{2 m^{\ast}\omega_0^2}\nonumber\\
\pm \sqrt{(\alpha^2 + \beta^2) k^2 + 4 \alpha \beta k_x k_y},
\end{eqnarray}
which is a particular form of Eq. (\ref{approx-energy}) written at
$B=0$, since Eq. (\ref{approx-energy}) is exact in this limit. 


By expressing $b_n$ in Eq. (\ref{eq-b}) through $a_n, a_{n\pm 1}$
and substituting into Eq. (\ref{eq-a}) we get an equation for the
vector ${\bf a} = \{a_0, a_1, a_2, \dots \}$. An equation
for the vector ${\bf b} = \{b_0, b_1, b_2, \dots \}$ is obtained by the same
way; finally we get:
\begin{eqnarray}
\hat{\bf N} {\bf a} = 0 \label{matrixN}\\
\hat{\bf M} {\bf b} = 0
\label{matrixM}
\end{eqnarray}
$\hat{\bf N}$ and $\hat{\bf M}$ are square penthadiagonal matrices
of infinite order with non-zero entries $N_{i,j} \ne 0$ ($M_{i,j}
\ne 0$) only if $|i-j| \le 2$, and $\hat{\bf N} = (\hat{\bf
M})^{\ast}$. Apart from the non-zero main diagonal $N_{n,n}$, the
matrix $\hat{\bf N}$ contains the first two diagonals, $N_{n, n
\pm 1}$ and $N_{n, n \pm 2}$, above and below it, which are given
as
\begin{flushleft}
\begin{eqnarray}
\hspace{-5.0mm}&&N_{n,n} = \left(\frac{E}{\hbar \omega} - n -
\frac{1}{2}\right) - \frac{|c_0|^2}{\frac{E}{\hbar \omega} - n -
\frac{1}{2}} -{}\qquad \qquad
\nonumber\\
\hspace{-5.0mm}&&{} - \frac{2 n |c_1|^2}{\frac{E}{\hbar \omega} -
n + \frac{1}{2}}-\frac{2(n+1)|c_1|^2}{\frac{E}{\hbar \omega} - n -
\frac{3}{2}};\qquad \qquad
\label{Nnn}\\
\hspace{-5.0mm}&&N_{n,n-1} = - \sqrt{2 n} \left(\frac{c_1^{\ast}
c_0}{\frac{E}{\hbar \omega} - n - \frac{1}{2}} + \frac{c_0^{\ast}
c_1}{\frac{E}{\hbar \omega} - n + \frac{1}{2}}\right);\qquad
\label{Nnn-1}\\
\hspace{-5.0mm}&&N_{n,n+1} = - \sqrt{2 (n+1)}
\left(\frac{c_1^{\ast} c_0}{\frac{E}{\hbar \omega} - n -
\frac{1}{2}} + \frac{c_0^{\ast} c_1}{\frac{E}{\hbar \omega} - n -
\frac{3}{2}}\right);\qquad
\label{Nnn+1}\\
\hspace{-5.0mm}&&N_{n,n-2} = - \frac{2 \sqrt{n(n-1)}
~|c_1|^2}{\frac{E}{\hbar \omega} - n - \frac{1}{2}};\qquad \qquad
\label{Nnn-2}\\
\hspace{-5.0mm}&&N_{n,n+2} = - \frac{2 \sqrt{(n+1)(n+2)}
~|c_1|^2}{\frac{E}{\hbar \omega} - n - \frac{3}{2}}.\qquad \qquad
\label{Nnn+2}
\end{eqnarray}
\end{flushleft}
The energy spectrum has to be found from the secular equation, by
equating the determinant of the matrix $\hat{\bf N}$ to zero. The
infinite penthadiagonal matrix is truncated down to the first $n$
rows and $n$ columns, the roots of which can be found by numeric
methods, \cite{nao11a,nao11b}. 

\section{III. Spin current in equilibrium}

The equilibrium spin current in the previous activities 
\cite{rashba03,tokatly08,sonin07,sonin07b,esl05,bv08} has been studied for a pure
2D electron gas in the absence 
of an external magnetic field. This section is addressed to study the spin-current in
a quasi-2D electron gas with finite thickness in the
presence of an in-plane magnetic field and the gate potential. In order to write the
continuity equation for the charge density 
$\rho = e |\Psi(z)|^2 = |\psi_{\uparrow}|^2 +|\psi_{\downarrow}|^2$ 
and for the $\gamma$-component of the spin density 
$S_{\gamma} = (\hbar/2)(\Psi^{\dagger} \sigma_{\gamma}\Psi)$, the Schr\"odinger
equations (\ref{eq-up}) and (\ref{eq-down}) are multiplied to
their complex-conjugate components $\psi_{\uparrow}^{\ast}$ or
$\psi_{\downarrow}^{\ast}$, which yields
\begin{eqnarray}
\frac{\partial \rho}{\partial t} + {\bf \nabla} \cdot{\bf J} =
0\nonumber\\
\frac{\partial S_{\gamma}}{\partial t} + {\bf \nabla} \cdot{\bf
J}^{S_{\gamma}} = G_{\gamma},
\label{spin-current}
\end{eqnarray}
where ${\bf J}$ and ${\bf J}^{S_{\gamma}}$ are the charge- and the
spin-current, correspondingly. A violation of the spin
conservation in the system results in an additional source term
(torque) $G_{\gamma}$ \cite{sonin07} in the spin-balance equation.
The components of the charge- and spin currents read 
\begin{eqnarray}
\hspace{-5.0mm}&&J_j = -i\frac{e \hbar}{2 m^{\ast}}\big(\Psi^{\dagger}\nabla_j
\Psi - \nabla_j \Psi^{\dagger} \Psi\big) -\frac{e\alpha}{\hbar}
\Psi^{\dagger}({\bf \sigma}\times {\hat z}_0)_j \Psi 
\nonumber\\
\hspace{-5.0mm}&&{}+ \frac{e\beta}{\hbar} \Psi^{\dagger} {\tilde \sigma}_j \Psi -
\frac{e^2}{m^{\ast}c}\psi^{\dagger} A_j \Psi,
\label{chargecur}
\end{eqnarray}
and
\begin{eqnarray}
\hspace{-5.0mm}&&J_j^{S_{\gamma}} =
-i\frac{\hbar^2}{4m^{\ast}}[\Psi^{\dagger}\sigma_{\gamma}\nabla_j\Psi
- \nabla_j \Psi^{\dagger}\sigma_{\gamma}\Psi] -{}
\nonumber\\
\hspace{-5.0mm}&&{}-\frac{\alpha}{4}\{\Psi^{\dagger}[\sigma_{\gamma}({\bf
\sigma}\times {\hat z_0})_j + ({\bf \sigma}\times {\hat
z_0})_j\sigma_{\gamma}]\Psi\} +{} \nonumber\\
\hspace{-5.0mm}&&{} + \frac{\beta}{4}\{\Psi^{\dagger}[{\tilde \sigma}_j
\sigma_{\gamma} + \sigma_{\gamma} {\tilde \sigma}_j]\Psi\} -
\frac{e \hbar}{2 m^{\ast} c}A_j(\Psi^{\dagger} \sigma_{\gamma}
\Psi).
\label{spincur}
\end{eqnarray}
where ${\tilde {\bf \sigma}} = \{\sigma_x, - \sigma_y, 0\}$, ${\hat
z}_0$ is a unit vector, normal to the electron gas plane, and
$A_j$ is jth component of the vector-potential. It is easy to see that the
 structure of the contributions coming
from Rashba and Dresselhaus terms is similar to the
Hamiltonian form given by Eqs. (\ref{so-r}) and (\ref{so-d}). 
The magnetic field has a contribution to the charge current
as well as to the spin current, which consists with contribution
to the charge current \cite{landau} obtained by means of Hamilton
method in the absence of the SO interactions.

The equilibrium spin- and charge currents at T=0 are found by averaging 
Eqs. (\ref{chargecur}) and (\ref{spincur}) over the stationary states, given by the 
wave functions (\ref{wave-up}) and (\ref{wave-down}), and by integrating
over all occupied states, which yields for the charge current 
\begin{eqnarray}
\hspace{-5.0mm}&&\langle J_x \rangle = \frac{e\hbar}{m^{\ast}}\sum_{n=0}^{n_{m}}
\int \frac{d^2 k}{(2\pi)^2}k_x -  \frac{e\alpha}{\hbar} \langle \sigma_y \rangle + 
\frac{e \beta}{\hbar} \langle \sigma_x \rangle = \nonumber \\
\hspace{-5.0mm}&&{} e \sum_{n=0}^{n_m} \int \frac{d^2k}{(2\pi)^2}\left\{ \frac{\hbar k_x}{m^{\ast}} +
 \frac{\beta + i\alpha}{\hbar} a_n^{\ast} b_n + \frac{\beta - i\alpha}{\hbar} b_n^{\ast} a_n 
\right\},
\label{charge-current-x}\\
\hspace{-5.0mm}&&{}\langle J_y \rangle =e\bigg\{ \frac{\hbar}{m^{\ast}}\sum_{n=0}^{n_{m}}
\int \frac{d^2 k}{(2\pi)^2} k_y + \frac{\alpha}{\hbar} \langle\sigma_x \rangle - 
\frac{\beta}{\hbar} \langle \sigma_y \rangle + \omega_B \langle z \rangle \bigg\} \nonumber\\
\hspace{-5.0mm}&&{}= e \sum_{n=0}^{n_{m}} \int \frac{d^2 k}{(2\pi)^2} \bigg\{ \frac{\hbar k_y}{m^{\ast}} + 
\frac{\alpha + i\beta}{\hbar} a_n^{\ast} b_n + \frac{\alpha - i\beta}{\hbar} b_n^{\ast} a_n + 
\label{charge-current-y}\nonumber\\ 
\hspace{-5.0mm}&&{} +\omega_B z_0 + \omega_B a_B \sqrt{2(n+1)}~(a_{n+1}^{\ast} a_n + 
b_{n+1}^{\ast}b_n)\bigg\},
\end{eqnarray}
\begin{equation}
\langle J_z \rangle = 0
\label{charge-current}
\end{equation}

Note that the equations (\ref{charge-current-x})-(\ref{charge-current})  
for the charge current components can be
obtained according to $\langle J_i\rangle = e \langle v_i \rangle$ as well
by using the Heisenberg equation of motion $v_i = \frac{d r_i}{d
t} = \frac{i}{\hbar} [{\hat H},r_i]$.

The average values of the spin current components read as
\begin{eqnarray}
\hspace{-5.0mm}&&\langle J_x^{S_j}\rangle =\sum_{n=0}^{n_m} \int \frac{d^2 k}{(2\pi)^2} (a_n^{\ast}
b_n^{\ast})\bigg(\frac{\hbar k_x}{m^{\ast}} {\hat S}_j + \frac{\alpha}{2} \epsilon_{jx} + 
\frac{\beta}{2} \epsilon_{jy}\bigg) {a_n \choose b_n}\nonumber\\
\hspace{-5.0mm}&&{}=\int \frac{d^2 k}{(2\pi)^2} \left\{\frac{\hbar^2 k_x}
{2m^{\ast}}\langle \sigma_j \rangle_{{\bf k},n_m}+
\frac{\alpha}{2}\epsilon_{jx}+\frac{\beta}{2}\epsilon_{jy}\right\},
\label{spin-current-x}
\end{eqnarray}
\begin{eqnarray}
\hspace{-5.0mm}&&\langle J_y^{S_j}\rangle = \sum_{n=0}^{n_m} \int \frac{d^2 k}{(2\pi)^2}\bigg \{(a_n^{\ast} 
b_n^{\ast}) \bigg( \frac{\hbar k_y}{m^{\ast}}\frac{\omega_0^2}{\omega^2} +
\frac{e E_g \omega_B}{m^{\ast} \omega^2} \bigg){\hat S}_j +\nonumber\\
\hspace{-5.0mm}&&{}+\frac{\alpha}{2} \epsilon_{jy} +\frac{\beta}{2} \epsilon_{jx} + 
\omega_B a_B \sqrt{2(n+1)} (a_{n+1}^{\ast}
b_{n+1}^{\ast}) {\hat S}_j \bigg\} {a_n \choose b_n}\nonumber\\
\hspace{-5.0mm}&&{}= \int \frac{d^2 k}{(2\pi)^2}\bigg \{ 
\bigg( \frac{\hbar^2 k_y}{m^{\ast}}\frac{\omega_0^2}{\omega^2} +
\frac{e E_g \omega_B \hbar}{m^{\ast} \omega^2} \bigg) \langle  \sigma_j\rangle_{{\bf k},n_m} +\nonumber\\
\hspace{-5.0mm}&&{}+\frac{\alpha}{2} \epsilon_{jy} +\frac{\beta}{2} \epsilon_{jx} + 
\frac{a_B}{2}\omega_B \hbar \langle \sigma_j\rangle_{{\bf k},n_m}^{off}\bigg\}  
\label{spin-current-y}
\end{eqnarray}
and
\begin{eqnarray}
\hspace{-5.0mm}&&\langle J_{k_z}^{S_z} \rangle=i\frac{\hbar}{m^{\ast} a_B}\sum_{n=0}^{n_m}\int 
\frac{d^2 k}{(2\pi)^2} 
 \sqrt{2(n+1)}\times \nonumber\\
\hspace{-5.0mm}&&{}(a_{n+1}^{\ast} b_{n+1}^{\ast}){\hat S}_z {a_n\choose b_n}
=i\frac{\hbar^2}{2m^{\ast}a_B}\sum_{n_m}\int\frac{d^2 k}{(2\pi)^2}\langle \sigma_z \rangle_{{\bf k},n_m}, 
\label{spin-current-z}
\end{eqnarray}
where $j=x,y$ and $\langle J_z^{S_x}\rangle = \langle J_z^{S_y}\rangle =0$, also 
$\langle J_x^{S_z}\rangle = \langle J_y^{S_z}\rangle = 0$; $\epsilon_{ij}$ is
a 2D antisymmetric tensor with components $\epsilon_{xy}=-\epsilon_{yx}=1$. The 
expressions for $\langle \sigma_x \rangle_{{\bf k},n_m}$, $\langle \sigma_y \rangle_{{\bf k},n_m}$,
$\langle \sigma_z \rangle_{{\bf k},n_m}$ and their evident momentum dependences are 
calculated in Appendix.

It is evident that the normal to the electron gas component of 
$\langle {\bf J}^{S_z} \rangle$ in
Eq. (\ref{spin-current-z}) arises exclusively due to the
transverse confinement and the in-plane magnetic field. Nevertheless this term
does not accumulate a spin, an electron tunneling from $n$th to
$(n+1)$th level is accompanied, according to
Eq. (\ref{spin-current-z}), by reverse flow from $(n+1)$th to $n$th
level.

The spin-continuity equation (\ref{spin-current}) contains the
source-term $G_{\gamma}$ due to a violation of the spin
conservation, the components of which are given by the following
expressions
\begin{eqnarray}
\hspace{-5.0mm}&&G_j =- i \frac{\alpha}{2}\{
\Psi^{\dagger}[{\bf \sigma} \times ({\hat z}_0 \times {\bf
\nabla})]_j \Psi - [({\bf \nabla} \times {\hat z}_0) \times {\bf
\sigma}]_j \Psi^{\dagger}
\Psi\}-{}  \nonumber\\
\hspace{-5.0mm}&&{}- i\frac{\beta}{2} \{\Psi^{\dagger}
({\tilde{\bf \nabla}} \times {\bf \sigma})_j \Psi - ({\tilde {\bf
\nabla}} \times {\bf \sigma})_j \Psi^{\dagger} \Psi \} + \beta
\frac{e}{c}\Psi^{\dagger}({\bf A} \times {\bf \sigma}) \Psi +{}
 \nonumber\\
\hspace{-5.0mm}&&{}+ \frac{1}{2}g \mu_B \Psi^{\dagger}({\bf B}
\times {\bf \sigma}) \Psi + \alpha \frac{e}{c}
\Psi^{\dagger}\left(({\hat z}_0 \times {\bf A})\times {\bf
\sigma}\right)\Psi, \label{torque}
\end{eqnarray}
where ${\bf {\tilde \nabla}}=\{\nabla_x, - \nabla_y,0\}$ and
${\bf{\tilde \sigma}} =\{\sigma_x, - \sigma_y,0\}$. The averaging
of x- and y-components of the torque over the quantum-mechanic
states gives
\begin{equation}
\langle G_x \rangle = 0, \qquad {\rm and} \qquad \langle G_y
\rangle = 0,
\end{equation}
which is in consistence with a result in strictly 2D system
\cite{sonin07} in the absence of an external magnetic field.
Nevertheless z-component of the torque is not averaged to zero:
\begin{eqnarray}
\hspace{-10.0mm} &&\langle G_z \rangle =\sum_{n=0}^{n_m} \int \frac{d^2 k}{(2\pi)^2}
\big\{(\alpha k_x + \beta k_y + \frac{\beta}{\hbar} z_0 \omega_B m^{\ast})
(a_n^{\ast} b_n +\nonumber\\
&&+b_n^{\ast} a_n)  - i (\alpha k_y + \beta k_x +
\frac{1}{2}g\mu_B B +
\frac{\alpha}{\hbar} z_0 \omega_B m^{\ast}) (a_n^{\ast} b_n -\nonumber\\
\hspace{-10.0mm}&& - b_n^{\ast} a_n)- i \frac{\alpha}{\hbar} a_B \omega_B
m^{\ast} \sqrt{2(n+1)}~(a_{n+1}^{\ast} b_n - b_{n+1}^{\ast}
a_n) +\nonumber\\
&& + \frac{\beta}{\hbar} a_B \omega_B m^{\ast}
\sqrt{2(n+1)}~(a_{n+1}^{\ast} b_n + b_{n+1}^{\ast} a_n) \big\}.
\label{torque-z}
\end{eqnarray}
The integration at $T=0$ is taken over each momentum component $i=x,y$ 
and for both spin branches in each transverse-quantized subband up to the Fermi level, 
$-K^i_{n_{m},\pm} \le k_i \le K^i_{n_{m}, \pm}$.
In order to calculate the average values of the charge- and spin current 
components, given by Eqs. (\ref{charge-current})-(\ref{spin-current}),
evident expressions of $a_n$ and $b_n$ are required. 
The Fermi level is assumed to be set between $n_m$ and $n_m+1$ subbands, 
so that all levels up to $\{{\bf
K}_m, n_m\}$ are occupied with $a_n \ne 0,~~b_n \ne 0$ for $n \le n_m$ and $a_n=b_n=0$
for $n > n_m$. A simplest case, which takes into account the inter-subband mixing due to 
an interference between the SO interactions and in-plane magnetic field, is $n_m=1$. In this 
case equations (\ref{eq-a}) and (\ref{eq-b}) are simplified to the form, given by Eq. (\ref{append-1})
in Appendix. The analytical expressions for the energy spectrum in the first and second
transverse-quantized subbands are given according to Eqs. (\ref{energyn1})-(\ref{energy})  
\begin{eqnarray}
\hspace{-5.0mm}&&E_{\pm}^{(n)}=\frac{\hbar^2 k^2}{2 m^{\ast}}-
\frac{(eE_g-\hbar \omega_Bk_y)^2}{2m^{\ast}\omega^2}+
\hbar \omega +\lambda_n\frac{\hbar \omega}{2}\times \nonumber\\ 
\hspace{-5.0mm}&&{}\sqrt{1+4|c_0|^2+8|c_1|^2 \mp 4\sqrt{|c_0|^2+
2(c_0^{\ast}c_1+c_0c_1^{\ast})^2}},
\label{energy-n}
\end{eqnarray}
where $\lambda_n$ for $n=0,1$ indicates the sub-band index with $\lambda_0=-$ and 
$\lambda_1=+$, and the sign $\pm$ shows the spin-branch index.
In-plane momentum dependence of the energy spectrum, Eq. (\ref{energy-n}), 
is determined by the terms 
$|c_0|^2$ and $(c_0^{\ast}c_1+c_0c_1^{\ast})^2$ 
\begin{eqnarray}
\hspace{-5.0mm}&&|c_0|^2=\frac{1}{\hbar^2 \omega^2}\bigg\{ (\alpha^2+\beta^2) \bigg[ \bigg(k_y
\frac{\omega_0^2}{\omega^2}+
\frac{eE_g\omega_B}{\hbar \omega^2}\bigg)^2+k_x^2\bigg]+ \nonumber\\ 
\hspace{-5.0mm}&&{}4\alpha \beta k_x \bigg(k_y\frac{\omega_0^2}
{\omega^2}+\frac{eE_g\omega_B}{\hbar \omega^2}\bigg)+\nonumber\\
\hspace{-5.0mm}&&2\omega_z \hbar \bigg[\alpha \bigg( k_y\frac{\omega_0^2}
{\omega^2}+\frac{eE_g\omega_B}{\hbar \omega^2}\bigg)+\beta k_x\bigg] \bigg\},
\label{mod-c0}
\end{eqnarray}

\begin{eqnarray}
\hspace{-5.0mm}&&c_0^{\ast}c_1+c_0c_1^{\ast}=\frac{\omega_B}
{(\omega \hbar)^2}\sqrt{\frac{m^{\ast}}{\omega \hbar}}
\bigg\{(\alpha^2+\beta^2) \bigg(k_y\frac{\omega_0^2}{\omega^2}+
\frac{eE_g\omega_B}{\hbar \omega^2}\bigg)+\nonumber\\
\hspace{-5.0mm}&&{}+2\alpha \beta k_x +\alpha \omega_z \hbar \bigg\}
\end{eqnarray}
So, $|c_0|^2\sim O(\alpha^2, \beta^2)$, whereas  $(c_0^{\ast}c_1+c_0c_1^{\ast})^2\sim 
O(\alpha^4, \beta^4, \alpha^2 \beta^2)$ in the absence of Zeeman splitting. 
Therefore, expansion of Eq. (\ref{energy-n}) over small 
SO coupling constants up to quadratic in $\alpha, \beta$ terms yields
\begin{equation}
E_{\pm}^{(0)}\approx \frac{\hbar^2 k^2}{2 m^{\ast}}-\frac{(eE_g-\hbar \omega_Bk_y)^2}
{2m^{\ast}\omega^2}+ \frac{1}{2}\hbar \omega \pm |c_0| \hbar \omega;
\label{energy0-approx}
\end{equation}
\begin{equation}
E_{\pm}^{(1)}\approx \frac{\hbar^2 k^2}{2 m^{\ast}}-\frac{(eE_g-\hbar \omega_Bk_y)^2}
{2m^{\ast}\omega^2}+ \frac{3}{2}\hbar \omega \mp |c_0|\hbar \omega.
\label{energy1-approx}
\end{equation} 
The limit of integration over the occupied states can be found by fixing the Fermi
energy $E_F$ in Eq. (\ref{energy-n}) and solving this equation for  
the momentum. It is necessary to note that an interference between the gate 
electric field and the orbital effect of the in-plane magnetic field shifts
the Fermi surface along $k_y$ axis (see, Eqs. (\ref{energy0-approx}), 
(\ref{energy1-approx}) and (\ref{mod-c0})). Furthermore, Zeeman splitting makes the 
energy spectra asymmetric along both $k_x$ and $k_y$ axes. Therefore, the 
integrations over $k_x$ and $k_y$ 
have to be taken, generally speaking, over asymmetric intervals 
$-K^{' x}_{n_{m},\pm} \le k_x \le K^x_{n_{m}, \pm}$ and 
$-K^{' y}_{n_{m},\pm} \le k_y \le K^y_{n_{m}, \pm}$. 

The integration limit is calculated in Appendix by fixing the Fermi energy and  
transforming the momentum components
in Eqs. (\ref{energy0-approx}) and (\ref{energy1-approx}) into
polar coordinates, $k_x=k \cos \varphi$, $k_y = k \sin \varphi$. We express Eq. (\ref{kF}) 
for $k_{n,\pm}^F$ as $k_{n,\pm}^F=k_n^F  \pm \delta k$, where  
\begin{widetext}
\begin{eqnarray}
\hspace{-5.0mm}&&k_{n}^F = \frac{1}{\cos^2 \varphi + \frac{\omega_0^2}{\omega^2}\sin^2 \varphi}
\bigg\{-\frac{eE_g \omega_B}{\hbar \omega^2}\sin \varphi +
\bigg[\frac{2m^{\ast}}{\hbar^2}[E_F - \omega \hbar (n+1/2)](\cos^2 \varphi + 
\frac{\omega_0^2}{\omega^2}\sin^2 \varphi)+\nonumber\\
\hspace{-5.0mm}&&{} \frac{e^2E_g^2}{\omega^2 \hbar^2}+
\frac{m^{\ast 2}}{\hbar^4}\big[(\alpha^2+\beta^2)\left( \cos^2\varphi + 
\frac{\omega_0^4}{\omega^4}\sin^2\varphi \right) + 4 \alpha \beta \frac{\omega_0^2}{\omega^2} 
\sin \varphi \cos \varphi \big]\bigg]^{1/2}\bigg\},\\
\hspace{-5.0mm}&&{}\delta k = \frac{m^{\ast}}{\hbar^2(\cos^2 \varphi + \frac{\omega_0^2}{\omega^2}\sin^2 \varphi)}
\sqrt{(\alpha^2+\beta^2)\left( \cos^2\varphi + 
\frac{\omega_0^4}{\omega^4}\sin^2\varphi \right) + 4 \alpha \beta \frac{\omega_0^2}{\omega^2} 
\sin \varphi \cos \varphi}.
\end{eqnarray}
\end{widetext}  
for $n=0,1$. 

The average values of the Pauli spin-matrices $\langle \sigma_j \rangle_{{\bf k},n_m}, j=x,y,z$ 
are calculated in Appendix. By using the expressions (\ref{c0}), (\ref{c1}), (\ref{mod-c0}) 
and (\ref{c0-c1}) in 
Eqs. (\ref{sigma-x}), (\ref{sigma-y}) and (\ref{sigma-z}) in Appendix, one gets an explicit 
momentum-dependent expressions for the averaged Pauli matrices
\begin{eqnarray}
\langle \sigma_x \rangle_{k_{\pm},n}= \pm \lambda_n \frac{a k +b}{\sqrt{ck^2+dk+e}},
\label{av-sigma-x}\\
\langle \sigma_y \rangle_{k_{\pm},n}= \mp \lambda_n i\frac{\bar{a} k +\bar{b}}
{\sqrt{ck^2+dk+e}},
\label{av-sigma-y} \\
\langle \sigma_z \rangle_{k_{\pm},n}= \mp \lambda_n i\frac{\tilde{a} k +\tilde{b}}
{\sqrt{ck^2+dk+e}},
\label{av-sigma-z}
\end{eqnarray}
where $k$ is the momentum modulus $\{k_x,k_y\}=\{k \cos \phi, k \sin \phi\}$ in 
spherical-polar system and
\begin{eqnarray}
\hspace{-5.0mm}&&a=\alpha\frac{\omega_0^2}{\omega^2}\sin \phi+\beta \cos \phi,
\label{a}\\
\hspace{-5.0mm}&&b=\alpha \frac{eE_g\omega_B}{\hbar \omega^2}+\omega_z\hbar,
\label{b}\\
\hspace{-5.0mm}&&\bar{a}=\alpha\cos \phi + \beta \frac{\omega_0^2}{\omega^2}\sin \phi,
\label{a-bar}\\
\hspace{-5.0mm}&&\bar{b}=\beta \frac{eE_g\omega_B}{\hbar \omega^2},
\label{b-bar}\\
\hspace{-5.0mm}&&\tilde{a}=\frac{\omega_B}{\omega \hbar}\sqrt{\frac{m^{\ast}}{\omega \hbar}}
(\alpha^2-\beta^2)\cos \varphi,
\label{a-tilde}\\
\hspace{-5.0mm}&&\tilde{b}=-\frac{\omega_B}{\omega}\sqrt{\frac{m^{\ast}}{\omega \hbar}}
\beta \omega_z,
\label{b-tilde}\\
\hspace{-5.0mm}&&c= \left(\alpha \frac{\omega_0^2}{\omega^2}\sin \phi + \beta \cos \phi\right)^2 + 
\left(\beta \frac{\omega_0^2}{\omega^2} \sin \phi+\alpha \cos \phi\right)^2,
\label{c}\\
\hspace{-5.0mm}&&d=2\left(\frac{2eE_g\omega_B}{\hbar \omega^2}\right)\bigg[\alpha 
\left(\alpha \frac{\omega_0^2}{\omega^2}\sin \phi 
+ \beta \cos \phi \right)+\beta\bigg(\beta\frac{\omega^2}{\omega^2}\sin \phi \nonumber\\
\hspace{-5.0mm}&&+\alpha \cos \phi\bigg)\bigg] 
+ 2\omega_z \hbar \left(\alpha\frac{\omega_0^2}{\omega^2}\sin \phi +\beta \cos \phi \right),
\label{d}\\
\hspace{-5.0mm}&&e=\left(\alpha\frac{eE_g\omega_B}{\hbar \omega^2}+\omega_z \hbar \right)^2 + 
\left(\beta \frac{eE_g\omega_B}{\hbar \omega^2}\right)^2.
\label{e}
\end{eqnarray}

In-plane magnetic field induces an inter-sub-band coupling terms, $\langle \sigma_i \rangle_{{\bf k},n_m}^{off}$
with $i=x,y$, which give a contribution to the $y$-components of the 
spin current (\ref{spin-current-y}). These terms are calculated in Appendix.
By neglecting the small terms $[c_0^{\ast 2}c_1+c_0^2c_1^{\ast}+(c_1^{\ast}+c_1)|c_0|^2]$ and 
$[c_0^{\ast 2}c_1-c_0^2c_1^{\ast}+(c_1^{\ast}-c_1)|c_0|^2]$, Eqs. (\ref{sigmax-off}) 
and (\ref{sigmay-off}) in Appendix are approximated as
\begin{eqnarray}
\hspace{-5.0mm}&&\langle \sigma_x \rangle_{{\bf k},n_m}^{off} 
\approx \frac{c_1^{\ast} + c_1}{2(\tilde{\epsilon}-1)}
=\lambda_n(c_1^{\ast}+c_1)=\nonumber\\
&&=\lambda_n\frac{\alpha \omega_B}{\omega \hbar}\sqrt{\frac{m^{\ast}}
{\omega \hbar}},
\label{sigma-x-off}\\
\hspace{-5.0mm}&&\langle \sigma_y \rangle_{{\bf k},n_m}^{off} 
\approx -i\frac{c_1^{\ast} - c_1}{2(\tilde{\epsilon}-1)}
=-i\lambda_n(c_1^{\ast}-c_1)= \nonumber\\
&&=-\lambda_n\frac{\beta \omega_B}{\omega \hbar}\sqrt{\frac{m^{\ast}}
{\omega \hbar}}.
\label{sigma-y-off}
\end{eqnarray}
So, the inter-subband coupling terms $\langle \sigma_x \rangle_{{\bf k},n_m}^{off}$ and 
$\langle \sigma_y \rangle_{{\bf k},n_m}^{off}$ depend only on the sub-band index and do 
not depend on the spin-branch index.

In order to calculate the equilibrium spin-current components, Eqs. (\ref{spin-current-x})-
(\ref{spin-current-z}) are integrated firstly over the momentum modulus $k$ by taking 
into account Eqs. (\ref{av-sigma-x})-(\ref{sigma-y-off}). Routine calculations 
yield under this condition the 
following results for the equilibrium spin-current ${\bf J}^{S_x}= \{J_x^{S_x},J_y^{S_x},0\}$
\begin{widetext}
\begin{eqnarray}
&&\langle J_x^{S_x}\rangle=\frac{m^{\ast 2}\omega_0}{3 \pi \hbar^4 \omega}\beta (\beta^2-\alpha^2) +
\frac{1}{2\pi \beta}\left(\frac{eE_g\omega_B}{\omega_0^2\hbar}\right)^2
\bigg\{\frac{1}{2}(\alpha^2+
2\beta^2 -|\alpha^2-\beta^2|)+\frac{1}{8\alpha^2}\big[(\alpha^2-\beta^2)^2-\nonumber\\
&&|\alpha^2-\beta^2|
(\alpha^2+\beta^2)\big]+\frac{\alpha^2+\beta^2}{3}\left(1-\frac{|\alpha^2-\beta^2|}
{(\alpha^2-\beta^2)}\right)\bigg\}+
\frac{\omega_z}{4\pi\alpha \beta}\frac{eE_g\omega_B}
{\omega_0^2}\bigg\{\alpha^2+\beta^2-|\alpha^2-\beta^2|+\nonumber\\
&&\frac{7\alpha^2}{3}\left(1-
\frac{|\alpha^2-\beta^2|}{(\alpha^2-\beta^2)}\right)\bigg\}+\frac{5\omega_z^2\hbar^2}{12\pi\beta}
\left\{1-\frac{|\alpha^2-\beta^2|}{(\alpha^2-\beta^2)}-\frac{\alpha^2 + \beta^2 -|\alpha^2-\beta^2|}{\alpha^2}\right\},
\label{Jx-Sx}\\
&&\langle J_y^{S_x}\rangle=\frac{m^{\ast 2}\omega_0}{3 \pi \hbar^4 \omega}\alpha \big[(\alpha^2-\beta^2)
+\frac{3\omega_B^2}{\omega_0^2}(\alpha^2+\beta^2)\big]+
\frac{\alpha m^{\ast}\omega_B^2}{\pi \hbar^2 \omega \omega_0}
\left(E_F-\frac{3}{2}\omega\hbar \right)  -
\frac{1}{48\pi \alpha\beta^2}\left(\frac{eE_g\omega_B}{\omega_0^2\hbar}\right)^2
\bigg\{3\alpha^4-\nonumber\\
&&-17\beta^4 - 62\alpha^2\beta^2 +3(3\beta^2-\alpha^2)|\alpha^2-\beta^2|-
8\beta^2\frac{(\alpha^4-\beta^4)}{|\alpha^2-\beta^2|}\bigg\} - \frac{\omega_z}
{12\pi\alpha^2 \beta^2}\frac{eE_g\omega_B}{\omega_0^2} \bigg\{3\alpha^4-3\beta^4-\nonumber\\
&&-3(\alpha^2-\beta^2)|\alpha^2-\beta^2| -6\alpha^2 \beta^2
\frac{|\alpha^2-\beta^2|}{(\alpha^2-\beta^2)}\bigg\}+\frac{5\omega_z^2\hbar^2}{12\pi\alpha}
\left\{1+\frac{|\alpha^2-\beta^2|}{(\alpha^2-\beta^2)}-\frac{\alpha^2+\beta^2-
|\alpha^2-\beta^2|}{\beta^2}\right\},
\label{Jy-Sx}
\end{eqnarray}
as well as for ${\bf J}^{S_y}=\{J_x^{S_y},J_y^{S_y},0\}$
\begin{eqnarray}
\hspace{-5.0mm}&&\langle J_x^{S_y}\rangle=\frac{m^{\ast 2}\omega_0}
{3 \pi \hbar^4 \omega}\alpha (\beta^2-\alpha^2) +
\frac{1}{48\pi \alpha\beta^2}\left(\frac{eE_g\omega_B}{\omega_0^2\hbar}\right)^2
\bigg\{3(\alpha^2+5\beta^2)|\alpha^2 - \beta^2| -(\alpha^2+
\beta^2)(3\alpha^2+23\beta^2)-\nonumber\\
\hspace{-5.0mm}&&{}-8\beta^2(\alpha^2+\beta^2)\frac{|\alpha^2-\beta^2|}
{(\alpha^2-\beta^2)}\bigg\}
-\frac{\omega_z}{24\pi\alpha^2 \beta^2}\frac{eE_g\omega_B}{\omega_0^2}
\bigg\{(\alpha^2-\beta^2)|\alpha^2-\beta^2|-(\alpha^4 -\beta^4)  +\nonumber\\
\hspace{-5.0mm}&&{}+2\alpha^2\beta^2
\frac{|\alpha^2-\beta^2|}{(\alpha^2-\beta^2)}\bigg\}-\frac{\omega_z^2\hbar^2}{6\pi\alpha}
\left\{1+ \frac{|\alpha^2-\beta^2|}{(\alpha^2-\beta^2)}-\frac{\alpha^2+
\beta^2-|\alpha^2-\beta^2|}{\beta^2}\right\},
\label{Jx-Sy}\\
\hspace{-5.0mm}&&\langle J_y^{S_y}\rangle=\frac{m^{\ast 2}\omega_0}
{3 \pi \hbar^4 \omega}\beta \left[(\alpha^2-\beta^2)
-\frac{3\omega_B^2}{\omega_0^2} (\alpha^2+\beta^2)\right]-
\frac{\beta m^{\ast}\omega_B^2}{\pi\hbar^2\omega \omega_0}\left(E_F-\frac{3}{2}
\omega \hbar \right) +
\frac{1}{48\pi \alpha^2\beta}\left(\frac{eE_g\omega_B}{\omega_0^2\hbar}\right)^2
\bigg\{ 3\alpha^4+\nonumber\\
\hspace{-5.0mm}&&{}+3\beta^4 -3(\alpha^2+\beta^2)|\alpha^2-\beta^2|-
8\alpha^2(\alpha^2+\beta^2)
\left(1-\frac{|\alpha^2-\beta^2|}{(\alpha^2-\beta^2)}\right)\bigg\} 
+ \frac{\omega_z}{24\pi\alpha \beta}\frac{eE_g\omega_B}{\omega_0^2} 
\bigg\{-5\alpha^2-3\beta^2-\nonumber\\
\hspace{-5.0mm}&&{}-3|\alpha^2-\beta^2| +2(4\alpha^2-3\beta^2)
\frac{|\alpha^2-\beta^2|}{(\alpha^2-\beta^2)}\bigg\}-
\frac{\omega_z^2\hbar^2}{6\pi\beta}
\left\{1-\frac{|\alpha^2-\beta^2|}{(\alpha^2-\beta^2)}-\frac{\alpha^2+\beta^2-
|\alpha^2-\beta^2|}{\alpha^2}\right\},
\label{Jy-Sy}
\end{eqnarray}
The non-zero component of $S_z$ spin-current, ${\bf J}^{S_z}=\{0,0,J_z^{S_z}\}$, induced 
by the in-plane magnetic field, can be presented as
\begin{eqnarray}
&&\langle J_z^{S_z}\rangle=\frac{m^{\ast}\omega_B(\alpha^2-\beta^2)}{4\pi\alpha \beta \omega \hbar^2}
\left\{(\alpha^2+\beta^2-|\alpha^2-\beta^2|)\left(\frac{eE_g\omega_B}{\hbar \omega^2}\right)
+\alpha \omega_z\hbar \left(1-\frac{|\alpha^2-\beta^2|}{\alpha^2-\beta^2}\right)\right\} +
\frac{m^{\ast}\beta\omega_B \omega_z}{\pi \omega_0 \hbar}.
\label{Jz-Sz}
\end{eqnarray}
The above expressions for the persistent spin current are simplified considerably as 
the gate voltage and Zeeman splitting vanish , $E_g=\omega_z=0$,
\begin{eqnarray}
&&{\bf J}^{S_x}=\left\{\frac{m^{\ast 2}\omega_0}{3\pi \hbar^4\omega}\beta (\beta^2-\alpha^2),~~
\frac{m^{\ast 2}\omega_0}{3\pi \hbar^4\omega}\alpha\left[(\alpha^2-\beta^2)+
\frac{3\omega_B^2}{\omega_0^2}(\alpha^2+\beta^2)\right]+\frac{\alpha m^{\ast}\omega_B^2}
{\pi \hbar^2 \omega \omega_0} \left(E_F-\frac{3}{2}\omega \hbar \right) ,~~0 \right\}
\label{J-Sx-E=0}\\
&&{\bf J}^{S_y}=\left\{\frac{m^{\ast 2}\omega_0}{3\pi \hbar^4\omega}\alpha (\beta^2-\alpha^2),~~
\frac{m^{\ast 2}\omega_0}{3\pi \hbar^4\omega}\beta \left[(\alpha^2-\beta^2) -\frac{3\omega_B^2}{\omega_0^2}(\alpha^2+\beta^2)\right]- \frac{\beta m^{\ast}\omega_B^2}
{\pi \hbar^2 \omega \omega_0}\left(E_F-\frac{3}{2}\omega \hbar\right),~~0 \right\};\\
&&{\bf J}^{S_z}=\left\{0,0,0\right\}.
\end{eqnarray}
\end{widetext}
Although the spin-current components turn to vanish at $\alpha=\beta$ in the absence of the 
magnetic field, the latter destroys this symmetry and yields a finite spin current as 
$\alpha \rightarrow \beta$. 
Indeed, the expressions (\ref{Jx-Sx})-(\ref{Jz-Sz}) yield the well-known results of 
Refs. \cite{esl05,bv08} for spin current in a $2D$ electron gas in the absence of the 
magnetic field
\begin{eqnarray}
{\bf J}^{S_x}=\left\{\frac{m^{\ast 2}}{3\pi \hbar^4}\beta (\beta^2-\alpha^2),~~
\frac{m^{\ast 2}}{3\pi \hbar^4}\alpha (\alpha^2-\beta^2),~~0 \right\};
\label{J-Sx-B=0}\\
{\bf J}^{S_y}=\left\{\frac{m^{\ast 2}}{3\pi \hbar^4}\alpha (\beta^2-\alpha^2),~~
\frac{m^{\ast 2}}{3\pi \hbar^4}\beta (\alpha^2-\beta^2),~~0 \right\},
\label{J-Sy-B=0}\\
{\bf J}^{S_z}=\left\{0,~~0,~~0\right\}.
\label{J-Sz-B=0}
\end{eqnarray}
It is easy to check that the diagonal components of the spin current vanish, $\langle J_x^{S_x}\rangle=
\langle J_y^{S_y}\rangle=\langle J_z^{S_z}\rangle=0$ as $\beta \rightarrow 0$ 
for arbitrary $\alpha \neq 0$. Nevertheless
\begin{widetext}
\begin{eqnarray}
&&\langle J_y^{S_x}\rangle = \frac{m^{\ast 2}\omega_0\alpha^3}{3\pi\hbar^4\omega}
\left(1+\frac{3\omega_B^2}{\omega^2}\right)+\frac{\alpha m^{\ast}\omega_B^2}{\pi \hbar^2\omega \omega_0}
\left(E_F-\frac{3}{2}\omega \hbar \right)+\frac{29\alpha}{24\pi}\left(\frac{eE_g\omega_B}
{\hbar \omega_0^2}\right)^2;\\
&& \langle J_x^{S_y}\rangle =-\frac{m^{\ast 2}\omega_0 \alpha^3}{3\pi\hbar^4 \omega}-
\frac{11\alpha}{24\pi}\left(\frac{eE_g\omega_B}{\hbar \omega_0^2}\right)^2,
\end{eqnarray}
under these conditions.
On the other hand the diagonal spin-current components are nonzero for $\alpha \rightarrow 0$ 
and $\beta \neq 0$
\begin{eqnarray}
\hspace{-5.0mm}&&\langle {\bf J}^{S_x}\rangle=\left\{\frac{m^{\ast 2}
\omega_0 \beta^3}{3\pi\hbar^4\omega}+\frac{11\beta}{24\pi}
\left(\frac{eE_g\omega_B}{\hbar\omega_0^2}\right)^2;~0;~0\right\},\\
\hspace{-5.0mm}&&{}\langle {\bf J}^{S_y}\rangle=\left\{0;
~-\frac{m^{\ast 2}\omega_0\beta^3}{3\pi\hbar^4\omega}
\left(1+\frac{3\omega_B^2}{\omega^2}\right)-\frac{\beta m^{\ast}\omega_B^2}
{\pi\hbar^2\omega_0\omega}\left(E_F-\frac{3}{2}\omega\hbar\right)-\frac{\beta}{3\pi}
\left(\frac{eE_g\omega_B}{\hbar\omega_0^2}\right)^2;~0\right\},\\
\hspace{-5.0mm}&&{}\langle {\bf J}^{S_z}\rangle = \left \{0;~0;~\frac{m^{\ast}\beta \omega_B \omega_z}
{2\pi \omega_0 \hbar}\right\}.
\end{eqnarray}
\end{widetext} 

New contributions to the pure 2D ($\sim \alpha^3, \beta^3$) spin-current expressions in 
Eqs. (\ref{Jx-Sx})-(\ref{Jy-Sy}) are caused by electron-transfer mechanism between nearest-neighbor 
transverse-quantized sub-bands, induced by the in-plane magnetic field, and they
vanish, consequently, with magnetic field. These terms are proportional either to the gate electric field 
or to Zeeman splitting too. Indeed, the gate electric field changes, on the one hand, 
the electronic energy spectrum, and shifts, on the other hand, the center of an magnetic
orbit $z_0$ along $z$-axis. The coefficient $c_0$, which has a physical meaning of 
probability amplitude
for an electron transition from one spin-polarized branch to other one in the same 
transverse-quantized subband $n$, according to Eqs. (\ref{eq-a}) and (\ref{eq-b}), 
parametrically depends on $E_g$ and $\omega_z$. The new terms in the spin-current 
expressions depend linearly on the SO coupling constants in the simplest case if one of the SO
coupling constant is zero. Therefore, these terms may dominate over the
pure 2D terms for some values of $\alpha, ~\beta,~E_g$ and magnetic field.   

\section{IV. Conclusions}

In this paper the equilibrium spin current is calculated for a quasi-2D
electron gas with finite thickness under in-plane magnetic field in the presence 
of Rashba- and Dresselhaus spin-orbit interactions. Note that the problem has been
solved for 2D electron gas in Refs. \cite{rashba03,tokatly08,sonin07,sonin07b,esl05,bv08} 
in the absence of the magnetic field. 
Our calculations show that the in-plane magnetic field generates out-of-plane spin current, 
which appears
exclusively due to sub-band mixing by means of in-plane magnetic field. Although the equilibrium
spin current vanishes at $\alpha = \beta$ in the absence of the magnetic field 
(see, Refs. \cite{esl05,bv08}),
in- plane magnetic field destroys this symmetry, yielding non-zero persistent spin-current at 
$\alpha=\beta$. The magnetic field strongly changes ${\bf J}^{S_x}$ and ${\bf J}^{S_y}$ 
spin-current 
components, and contributes new terms to them.
 
\section{Acknowledgment}
This research was supported by the DFG under grant Op28/8-1 and by the governmental
grant of Azerbaijan Republic under grant EIF-2010-1(1)-40/01-22.

\section{Appendix}

Equations (\ref{eq-a}) and (\ref{eq-b}) are reduced for $n=1$ into the following form
\begin{eqnarray}
(\tilde{\epsilon}-3/2)~a_1 -c_0~b_1-\sqrt{2}c_1~b_0=0\nonumber\\
(\tilde{\epsilon}-3/2)~b_1-c_0^{\ast}~a_1-\sqrt{2}c_1^{\ast}~a_0=0\nonumber\\
(\tilde{\epsilon}-1/2)~a_0-c_0~b_0-\sqrt{2} c_1~b_1=0\nonumber\\
(\tilde{\epsilon}-1/2)~b_0-c_0^{\ast}~a_0-\sqrt{2}c_1^{\ast}~a_1=0,
\label{append-1}
\end{eqnarray}
where $\tilde{\epsilon}=\tilde{E}/(\hbar~\omega)$, and the higher order terms 
$a_2$ and $b_2$ are neglected, since the levels $n=0$ and $n=1$ only are filled. 
Excluding $a_1$ and $b_1$ from these equations one gets 
\begin{widetext}
\begin{eqnarray}
\bigg\{-\frac{c_0^{\ast}}{c_1^{\ast}}(\tilde{\epsilon}-\frac{3}{2})-\frac{c_0}{c_1}
(\tilde{\epsilon}-\frac{1}{2})\bigg\}a_0+\bigg\{\frac{(\tilde{\epsilon}-3/2)
(\tilde{\epsilon}-1/2)}{c_1^{\ast}}+\frac{c_0^2}{c_1}-2c_1\bigg\}b_0 = 0,
\label{a0}\\
\bigg\{\frac{(\tilde{\epsilon}-3/2)
(\tilde{\epsilon}-1/2)}{c_1}+\frac{c_0^{\ast 2}}{c_1^{\ast}}-2c_1^{\ast}\bigg\}a_0+
\bigg\{-\frac{c_0}{c_1}(\tilde{\epsilon}-\frac{3}{2})-\frac{c_0^{\ast}}{c_1^{\ast}}
(\tilde{\epsilon}-\frac{1}{2})\bigg\}b_0 =0.
\label{b0}
\end{eqnarray}

Eqs. (\ref{append-1}) yield a system of equations for $a_1$ and $b_1$ too, which differs from 
Eqs. (\ref{a0}) and (\ref{b0}) by interchanging the coefficients in the front of $a_0$ in 
Eq. (\ref{a0}) and of $b_0$ in Eq. (\ref{b0}). Both system of equations results in the 
following expression for the energy spectrum
\begin{eqnarray}
\big[(\tilde{\epsilon}-3/2)(\tilde{\epsilon}-1/2)- 2|c_1|^2\big]^2-|c_0|^2\big[(\tilde{\epsilon}-3/2)^2+
(\tilde{\epsilon}-1/2)^2\big]-2(c_0^{\ast 2}c_1^2+c_0^2c_1^{\ast 2})+|c_0|^4=0.
\label{energyn1}
\end{eqnarray}
\end{widetext}
By using the relation 
$(\tilde{\epsilon}-3/2)^2+(\tilde{\epsilon}-1/2)^2=2(\tilde{\epsilon}-3/2)
(\tilde{\epsilon}-1/2)+1$ in the second term of Eq. (\ref{energyn1}) one gets a 
quadratic equation for 
$z=(\tilde{\epsilon}-3/2)(\tilde{\epsilon}-1/2)$
\begin{eqnarray}
&&z^2-2(|c_0|^2+2|c_1|^2)z-|c_0|^2+|c_0|^4+4|c_1|^4-\nonumber\\
&&-2(c_0^{\ast 2}c_1^2+c_0^2c_1^{\ast 2})=0
\label{energy-z}
\end{eqnarray}
with the following solutions
\begin{equation}
z=|c_0|^2+2|c_1|^2 \pm \sqrt{|c_0|^2+2(c_0^{\ast}c_1+c_0c_1^{\ast})^2}.
\label{z}
\end{equation}
Solution of Eq. (\ref{z}) for the dimensionless energy $\tilde{\epsilon}$ is expressed as
\begin{equation}
\hspace{-5.0mm}\tilde{\epsilon}_{\pm}^{(n)}=1 + \frac{\lambda_n}{2}\sqrt{1+4|c_0|^2+8|c_1|^2 \mp 4\sqrt{|c_0|^2+
2(c_0^{\ast}c_1+c_0c_1^{\ast})^2}},
\label{energy}
\end{equation}
where $\lambda_n=\pm$ indicates the first ($\lambda_0=-$) and second ($\lambda_1=+$) energy 
subbands, and $\mp$ assigns two spin-polarized branches in each energy subband. 
Note that the coefficients $c_0$ and $c_1$ are given by Eqs. (\ref{c0}) 
and (\ref{c1}), furthermore $c_0$ only depends on the in-plane momentum components 
$\{k_x,k_y\}$.

The coefficients $a_n$ and $b_n$ are complex parameters. Solutions of Eqs. (\ref{a0}) 
and (\ref{b0}) with the normalization condition yield for the modulus $|a_0|=|b_0|$
\nopagebreak
\begin{widetext}
\begin{eqnarray}
|a_0|^2=\frac{\big(\tilde{\epsilon}-\frac{3}{2}\big)\big(\tilde{\epsilon}-\frac{1}{2}\big)
\big[\big(\tilde{\epsilon}-\frac{3}{2}\big)\big(\tilde{\epsilon}-\frac{1}{2}\big) 
-|c_0|^2-2|c_1|^2\big]^2-|c_0|^2\big[\big(\tilde{\epsilon}-\frac{3}{2}\big)
\big(\tilde{\epsilon}-\frac{1}{2}\big)- 2|c_1|^2\big]}
{4\big(\tilde{\epsilon}-1\big) \big[ \big(\tilde{\epsilon}-\frac{3}{2}\big)
\big(\tilde{\epsilon}-\frac{1}{2}\big) - |c_0|^2 -2|c_1|^2\big] \big[
\big(\tilde{\epsilon}-\frac{3}{2}\big)\big(\tilde{\epsilon}-\frac{1}{2}\big)-
\big(\tilde{\epsilon}-\frac{1}{2}\big)\big(|c_0|^2+2|c_1|^2\big)+|c_0|^2 \big]}.
\end{eqnarray}
\end{widetext}
In the absence of the magnetic field $B \to 0$ or $c_1 \to 0$ inter-subband
coupling disappears, and Eq. (\ref{z}) is reduced to the form
\begin{equation}
\big[(\tilde{\epsilon}-3/2)^2-|c_0|^2\big]\big[\tilde{\epsilon}-1/2]^2-|c_0|^2\big]=0
\end{equation}
The expression for $|a_0|^2$ is simplified as $|a_0|^2=1/2$ under this condition.

The symmetry relations $a_n=e^{i\theta} b_n^{\ast}$ and $b_n=e^{i\theta}a_n^{\ast}$ for
the complex coefficients $a_n=|a_n|e^{i \phi_n^a}$ and $b_n=|b_n|e^{i\phi_n^b}$ imply that a
total phase $\theta= \phi_n^a + \phi_n^b$ is undefined parameter, whereas the relative phase
$\phi_n^a-\phi_n^b$ for $n=0$ and $n=1$ can be defined from the relations (\ref{a0}) and (\ref{b0})
\begin{eqnarray}
\frac{b_n}{a_n}=\pm \exp\{i(\phi_n^b-\phi_n^a)\},\qquad n=0,~1;\\
\phi_0^b-\phi_0^a= arg \bigg[\frac{c_0^{\ast}}{c_1^{\ast}}(\tilde{\epsilon}-\frac{3}{2})+
\frac{c_0}{c_1}(\tilde{\epsilon}-\frac{1}{2})\bigg] -\nonumber\\ 
arg \bigg[\frac{(\tilde{\epsilon}-3/2)
(\tilde{\epsilon}-1/2)}{c_1^{\ast}}+\frac{c_0^2}{c_1}-2c_1\bigg];\\
\phi_1^b-\phi_1^a= arg \bigg[\frac{c_0}{c_1}(\tilde{\epsilon}-\frac{3}{2})+
\frac{c_0^{\ast}}{c_1^{\ast}}(\tilde{\epsilon}-\frac{1}{2})\bigg] -\nonumber\\ 
arg \bigg[\frac{(\tilde{\epsilon}-3/2)
(\tilde{\epsilon}-1/2)}{c_1^{\ast}}+\frac{c_0^2}{c_1}-2c_1\bigg].
\end{eqnarray}
The wave function can be expressed as
\begin{eqnarray}
\Psi(x,y,z)=\exp \{i(k_xx+k_yy)-(z-z_0)^2/2a_B^2\}\nonumber\\
\sum_{n=0}^{\infty}\frac{H_n((z-z_0)/a_B)}
{\sqrt{a_B\sqrt \pi~2^nn!}}a_n{1 \choose \pm e^{i(\phi_n^b-\phi_n^a)}},
\end{eqnarray}
where the signs $\pm$ correspond to two different spin-polarized branches.

In order to estimate the equilibrium charge- and spin- currents we have to calculate 
the mean values of the Pauli matrices $\langle \sigma_i \rangle_{{\bf k},n_m}$ 
in the eigenstates given by Eqs. (\ref{wave-up}) and (\ref{wave-down}) up to the Fermi 
level $n=n_m$, and integrate the results over $\{k_x,k_y\}$ components of the in-plane 
momentum vector, $-K^i_{n_{m},\pm} \le k_i \le K^i_{n_{m}, \pm}$. Routine calculations yield
the following results for $\langle \sigma_x \rangle_{{\bf k},n_m}$
\begin{widetext}
\begin{eqnarray}
&&\langle \sigma_x \rangle_{{\bf k},n_m}=\sum_{n=0}^{n_m=1}(a_n^{\ast}
b_n^{\ast}) \sigma_x {a_n \choose b_n}=a_0^{\ast}b_0+b_0^{\ast}a_0+a_1^{\ast}b_1+b_1^{\ast}a_1=\nonumber\\
&&=\frac{(c_0^{\ast}+c_0)
\big[\big(\tilde{\epsilon}-\frac{3}{2}\big)^2+
\big(\tilde{\epsilon}-\frac{1}{2}\big)^2-2|c_0|^2\big]+
4(c_0^{\ast}c_1^2+c_0c_1^{\ast 2})}
{4\big(\tilde{\epsilon}-1\big) \big[ \big(\tilde{\epsilon}-\frac{3}{2}\big)
\big(\tilde{\epsilon}-\frac{1}{2}\big) - |c_0|^2 -2|c_1|^2\big]},
\label{sigmax-av}
\end{eqnarray}
and for $\langle \sigma_y \rangle_{{\bf k},n_m}$ 
\begin{eqnarray}
\hspace{-5.0mm}&&\langle \sigma_y \rangle_{{\bf k},n_m}=\sum_{n=0}^{n_m=1}(a_n^{\ast}
b_n^{\ast}) \sigma_y {a_n \choose b_n}=-i(a_0^{\ast}b_0-b_0^{\ast}a_0+a_1^{\ast}b_1-b_1^{\ast}a_1)=\nonumber\\
&&=\mp i \lambda_n\frac{(c_0^{\ast}-c_0)
\big[\big(\tilde{\epsilon}-\frac{3}{2}\big)^2+
\big(\tilde{\epsilon}-\frac{1}{2}\big)^2-2|c_0|^2\big]-
4(c_0^{\ast}c_1^2-c_0c_1^{\ast 2})}
{4\big(\tilde{\epsilon}-1\big) \big[ \big(\tilde{\epsilon}-\frac{3}{2}\big)
\big(\tilde{\epsilon}-\frac{1}{2}\big) - |c_0|^2 -2|c_1|^2\big]}.
\label{sigmay-av}
\end{eqnarray}
\end{widetext}
By neglecting the small terms $4(c_0^{\ast}c_1^2+c_0c_1^{\ast 2})$ and 
$4(c_0^{\ast}c_1^2-c_0c_1^{\ast 2})$  in the numerators of Eqs. (\ref{sigmax-av}) and 
(\ref{sigmay-av}), correspondingly, and by using the expressions
(\ref{energy-z}) and (\ref{energy}) for the energy spectrum, one gets for 
$\langle \sigma_x \rangle_{k_{\pm},n}$
\begin{equation}
\langle \sigma_x\rangle_{k_{\pm},n}= \pm \lambda_n \frac{c_0^{\ast}+c_0}{2|c_0|},
\label{sigma-x}
\end{equation}
and for $\langle \sigma_y\rangle_{k_{\pm},n}$
\begin{equation}
\langle \sigma_y\rangle_{k_{\pm},n}= \mp \lambda_n i \frac{c_0^{\ast}-c_0}{2|c_0|}.
\label{sigma-y}
\end{equation}

$y$-components of the spin-current contain in addition an inter-subband coupling terms 
$\langle \sigma_i \rangle_{{\bf k},n_m}^{off}$
\begin{widetext}
\begin{eqnarray}
&&\langle \sigma_x \rangle_{{\bf k},n_m}^{off}=\sum_{n=0}^{n_m=1}\sqrt{2(n+1)}(a_{n+1}^{\ast}
b_{n+1}^{\ast}) \sigma_x {a_n \choose b_n}=\sqrt{2}(a_1^{\ast}b_0+b_1^{\ast}a_0)=\nonumber\\
&&\frac{(c_1^{\ast}+c_1)
\big[\big(\tilde{\epsilon}-\frac{3}{2}\big)
\big(\tilde{\epsilon}-\frac{1}{2}\big)-2|c_1|^2\big]+
(c_0^{\ast 2}c_1+c_0^2c_1^{\ast})}
{2\big(\tilde{\epsilon}-1\big) \big[ \big(\tilde{\epsilon}-\frac{3}{2}\big)
\big(\tilde{\epsilon}-\frac{1}{2}\big) - |c_0|^2 -2|c_1|^2\big]},\\
\label{sigmax-off}
\hspace{-5.0mm}&&\langle \sigma_y \rangle_{{\bf k},n_m}^{off}=\sum_{n=0}^{n_m=1}\sqrt{2(n+1)}
(a_{n+1}^{\ast}
b_{n+1}^{\ast}) \sigma_y {a_n \choose b_n}=-i(a_1^{\ast}b_0-b_1^{\ast}a_0)=
\nonumber\\
&&-i\frac{(c_1^{\ast}-c_1)
\big[\big(\tilde{\epsilon}-\frac{3}{2}\big) \big(\tilde{\epsilon}-\frac{1}{2}\big)
-2|c_1|^2\big]+(c_0^{\ast 2}c_1-c_0^2c_1^{\ast})}
{2\big(\tilde{\epsilon}-1\big) \big[ \big(\tilde{\epsilon}-\frac{3}{2}\big)
\big(\tilde{\epsilon}-\frac{1}{2}\big) - |c_0|^2 -2|c_1|^2\big]}.
\label{sigmay-off}
\end{eqnarray}

The transverse component of the spin current $\langle {\bf J}^{S_z}\rangle$ is proportional to 
$\langle \sigma_z \rangle_{{\bf k},n_m}$, which is given by a simple expression
\begin{eqnarray}
&&\langle \sigma_z \rangle_{{\bf k},n_m} = \sum_{n=0}^{n_m=1}\sqrt{2(n+1)}
(a_{n+1}^{\ast}
b_{n+1}^{\ast}) \sigma_z {a_n \choose b_n}=\sqrt{2}(a_1^{\ast}a_0-b_1^{\ast}b_0)=\nonumber\\
&&{} \frac{(c_0^{\ast}c_1-c_0c_1^{\ast})}
{2\big(\tilde{\epsilon}-1\big) \big[ \big(\tilde{\epsilon}-\frac{3}{2}\big)
\big(\tilde{\epsilon}-\frac{1}{2}\big) - |c_0|^2 -2|c_1|^2\big]}=\pm \lambda_n\frac{c_0^{\ast}c_1-
c_0c_1^{\ast}}{|c_0|}.
\label{sigma-z}
\end{eqnarray}
\end{widetext}
The momentum dependent factor $c_0^{\ast}c_1-c_0c_1^{\ast}$ in 
$\langle \sigma_z \rangle_{{\bf k},n_m}$ is given as
\begin{equation}
c_0^{\ast}c_1-c_0c_1^{\ast}=-i\frac{\omega_B}{(\omega \hbar)^2}
\sqrt{\frac{m^{\ast}}{\omega \hbar}}\left[(\alpha^2 - \beta^2)k_x-\beta \omega_z\right].
\label{c0-c1}
\end{equation}

In order to find the limit of integration in momentum space we transform the momentum components
$\{k_x,k_y \}$ into polar coordinates $k_x= k \cos \varphi$, $k_y = k \sin \varphi$ and 
write Eqs. (\ref{energy0-approx}) and (\ref{energy1-approx}) with fixed Fermi energy as
\begin{equation}
(\kappa^2-A_0)^2=A_1 \kappa^2 +A_2 \kappa +A_3,
\label{quartic}
\end{equation}
where $\kappa = k+\frac{eE_g\omega_B}{\hbar \omega^2}\frac{\sin \varphi}{\cos^2 \varphi + 
(\omega_0^2/\omega^2) \sin^2 \varphi}$ is a shifted momentum modulus, and
\begin{widetext}
\begin{eqnarray}
\hspace{-5.0mm}&&A_0=\frac{\big[E_F - \frac{\hbar \omega}{2}(n+1)\big](\cos^2\varphi + 
\frac{\omega_0^2}{\omega^2} \sin^2 \varphi) 
+ \frac{e^2 E_g^2}{2m^{\ast}\omega^2}}{\frac{\hbar^2}{2m^{\ast}}
(\cos^2 \varphi + \frac{\omega_0^2}{\omega^2} \sin^2 \varphi)^2};\\
\hspace{-5.0mm}&&{} A_1=\frac{(\alpha \cos \varphi + \beta \frac{\omega_0^2}{\omega^2} \sin \varphi)^2 + 
(\beta \cos \varphi + \alpha \frac{\omega_0^2}{\omega^2} \sin \varphi)^2}
{\frac{\hbar^4}{4m^{\ast 2}}(\cos^2 \varphi + \frac{\omega_0^2}{\omega^2} \sin^2 \varphi)^2};\\
\hspace{-5.0mm}&&{} A_2=\frac{8eE_gm^{\ast 2}\omega_B /(\hbar^5 \omega^2)\cos \varphi}{(\cos^2 \varphi 
+\frac{\omega_0^2}{\omega^2}\sin^2 \varphi)^3}\left\{2\alpha \beta (\cos^2 \varphi 
- \frac{\omega_0^2}{\omega^2}\sin^2 \varphi)-(\alpha^2+\beta^2)\frac{\omega_B^2}{\omega^2}
\sin \varphi \cos \varphi\right\}+\nonumber\\
&&\frac{8\omega_z m^{\ast 2}(\alpha \frac{\omega_0^2}{\omega^2}
\sin \varphi + \beta \cos \varphi)}{\hbar^4 (\cos^2 \varphi + 
\frac{\omega_0^2}{\omega^2}\sin^2 \varphi)^2}\\
\hspace{-5.0mm}&&{} A_3=\frac{\left(\frac{2eE_g m^{\ast}\omega_B}{\hbar^3 \omega^2}\right)^2 
\cos^2 \varphi(\alpha^2+\beta^2
-4\alpha \beta \sin \varphi \cos \varphi)}{(\cos^2 \varphi + 
\frac{\omega_0^2}{\omega^2}\sin^2 \varphi)^4}
+\frac{8eE_gm^{\ast 2}\omega_B \omega_z}{\hbar^5 \omega^2}\frac{\cos \varphi (\alpha \cos \varphi - 
\beta \sin \varphi)}{(\cos^2 \varphi + \frac{\omega_0^2}{\omega^2} \sin^2 \varphi)^4}+\nonumber\\
&&\frac{4m^{\ast 2}\omega_z^2}{\hbar^4(\cos^2 \varphi + \frac{\omega_0^2}{\omega^2}\sin^2 \varphi)^2}.
\label{A0-A4}
\end{eqnarray}
The parameter $A_0$ does not depend on small SO coupling constants $\alpha$ and $\beta$ and 
weakly depends on the gate electric field and in-plane magnetic field, whereas 
$A_1 \sim  O(\alpha^2, \beta^2)$, $A_2 \sim O(E_g \omega_B \alpha^2, E_g \omega_B \beta^2)$,
and $A_3 \sim O(E_g^2 \omega_B^2 \alpha^2, E_g^2 \omega_B^2 \beta^2)$.
By introducing an unknowing parameter $y$ Eq. (\ref{quartic}) can be rewritten as
\begin{equation}
(\kappa^2-A_0+y)^2=(A_1+2y)\kappa^2 +A_2\kappa +A_3-2yA_0+y^2
=(A_1+2y)\left[\kappa+\frac{A_2}{2(A_1+2y)}\right]^2+R(y).
\label{kappaF}
\end{equation}
The parameter $y$ is found under the condition 
\begin{eqnarray}
&&R(y)=0\quad {\rm or}\nonumber\\ 
&&y^3+\frac{A_1-4A_0}{2}y^2-(A_0A_1-A_3)y+\frac{A_3A_1-4A_2^2}{8}=0,
\end{eqnarray}
yielding
\begin{equation}
y=\frac{4A_0-A_1}{6}+\sqrt[3]{-\frac{Q}{2} \pm \sqrt{\frac{Q^2}{4}+\frac{P^3}{27}}}-
\frac{P}{3\sqrt[3]{-\frac{Q}{2}\pm \sqrt{\frac{Q^2}{4}+\frac{P^3}{27}}}},
\end{equation}
where
\begin{eqnarray}
P=-A_0A_1+A_3-\frac{(4A_0-A_1)^2}{12},\nonumber\\
Q=-\frac{(4A_0-A_1)^3}{108}+\frac{(A_0A_1-A_3)(4A_0-A_1)}{6}+\frac{4A_3A_1-A_2^2}{8}.
\end{eqnarray}
Expression for $y$ can be simplified to the form
\begin{equation}
y=2A_0 -\frac{A_3}{2A_0}.
\end{equation}
Finally, the Fermi momentum $k_{n,\pm}^F$ for each sub-band and spin-branch is found 
from Eq. (\ref{kappaF})
under the condition $R(y)=0$
\begin{eqnarray}
&&k^F_{n,\pm}= -\frac{eE_g\omega_B}{\hbar \omega^2}\frac{\sin \varphi}{\cos^2 \varphi +
\frac{\omega_0^2}{\omega^2} \sin^2 \varphi}+\sqrt{A_0+\frac{A_1}{4}-\frac{A_3}{4A_0}}\pm 
\sqrt{\frac{A_1}{4}+\frac{A_3}{4A_0}+\frac{A_2}{4\sqrt{A_0}}}.
\label{kF}
\end{eqnarray}
\end{widetext}
The equation (\ref{kF}) with Eqs. (\ref{A0-A4}) yields an evident expression for $k^F_{n,\pm}$.

\end{document}